\documentclass{m2an}

\usepackage{amsmath}
\usepackage{amssymb}

\newcommand{\E}{\mathrm{e}}
\newcommand{\B}{\mathcal{B}}
\newcommand{\Or}{\mathcal{O}}
\newcommand{\I}{\mathrm{i}}
\newcommand{\D}{\mathrm{d}}
\newcommand{\epsi}{\varepsilon}
\newcommand{\N}{\mathbb{N}}
\newcommand{\R}{\mathbb{R}}

\newcommand{\C}{\mathbb{C}}
\newcommand{\A}{{\cal A}}
\renewcommand{\H}{{\cal H}}

\newcommand{\e}{{\varepsilon }}
\def\d{{\partial}}

\def\({\left(}
\def\){\right)}
\def\<{\left\langle}
\def\>{\right\rangle}
\def\O{\mathcal O}

\newcommand{\ph}{\varphi}

\newcommand{\M}{\mathcal{M}}
\newcommand{\Hf}{\mathcal{H}_{\rm f}}

\newcommand{\m}{\ell}
\newcommand{\indexm}{= 1,\ldots, \ell}

\newcommand{\be}{\begin{equation}}
\newcommand{\ee}{\end{equation}}

\DeclareMathOperator{\Ran}{Ran}

\begin{document}

\title{The time-dependent Born-Oppenheimer approximation}
\author{Gianluca Panati}\address{Zentrum Mathematik, TU M\"unchen; \email{panati@ma.tum.de, spohn@ma.tum.de}}
\author{Herbert Spohn}\sameaddress{1}
\author{Stefan Teufel}\address{Mathematisches Institut, Universit\"at T\"ubingen;
\email{stefan.teufel@uni-tuebingen.de}}

\subjclass{81Q05, 81Q15, 81Q70.}

\keywords{Schr\"{odinger} equation, Born-Oppenheimer approximation, adiabatic methods, almost-invariant
subspace.}

\begin{abstract}
We explain why the conventional argument for deriving the time-dependent Born-Oppen\-heimer approximation is incomplete and review recent mathematical results, which clarify the situation and at the same time provide a systematic scheme for higher order corrections. We also present a new elementary derivation of the correct second-order time-dependent Born-Oppenheimer approximation and discuss as applications
the dynamics near a conical intersection of potential surfaces and reactive scattering.
\end{abstract}

\maketitle

\section{Introduction}

Through the discovery of Schr\"odinger the theoretical physics
community attained a powerful tool for
computing atomic spectra, either exactly or in
perturbation expansion. Born and Oppenheimer \cite{BoOp} immediately
strived for a more ambitious goal, namely to understand the
excitation spectrum of molecules on the basis of the new wave
mechanics. They accomplished to exploit the small electron/nucleus
mass ratio as an expansion parameter, which then leads to the {\em
static} Born-Oppenheimer approximation. Since then it has become a
standard and widely
used tool in quantum chemistry supported by a vast number of mathematical results, the classical ones being \cite{CDS,Ha1b,KMSW}.

Beyond excitation spectra and stationary scattering, dynamical processes gain
increasingly in interest. Examples are  chemical reactions or the decay of an excited state of
a molecule. Such problems are conveniently described within a
{\em time-dependent} version of the Born-Oppenheimer approximation, which
is the topic of this article. The starting point  is the observation
that the electronic energy at fixed position of the nuclei serves
as an effective potential between the nuclei. We call this the
zeroth order Born-Oppenheimer approximation. The resulting effective
Schr\"odinger equation can be used for both static and dynamic
purposes. Of course, the input is an electronic structure
calculation, which for the purpose of our article
we regard as given by other means.

While there are many physical and chemical properties of molecules
explained by the zeroth order Born-Oppen\-heimer approximation, there are cases
where higher order corrections are required. Famous examples are
the dynamical Jahn-Teller effect and the tunneling of singled out
nuclear degrees of freedom at a conical intersection of two Born-Oppenheimer energy surfaces. The first order Born-Oppenheimer approximation involves
geometric phases, which are of great interest also in other
domains of non-relativistic quantum mechanics \cite{BMKNZ}. Our plan is to
repeat in some detail the conventional argument for the first order
Born-Oppenheimer approximation. Only then the reader will
appreciate why a more systematic approach is in demand, which will be
the focus of our contribution.

Let us start from the molecular Hamiltonian
\begin{equation}\label{Hamiltonian1}
H_{\rm mol}= - \frac{\hbar^2}{2m_{\rm n}} \Delta_x - \frac{\hbar^2}{2m_{\rm e}}
\Delta_y + V(x, y)\,.
\end{equation}
 Here $x = \{ x_1, \ldots, x_K\}$ are the positions of the $K$ nuclei and
$y= \{ y_1, \ldots, y_N \} $ the positions of the $N$
electrons. The electrons have mass $m_{\rm e}$ and the nuclei have, for
notational simplicity, all the same mass $m_{\rm n}$. $V (x, y)$ is
the interaction potential, i.e.\ the Coulomb potential for
electrons and nuclei. For the mathematical results to be valid in the form we present them, one needs to slightly smear out the charge distribution of the nuclei. This is in line with the physical picture that nuclei are not pointlike but extended objects.  We transform \eqref{Hamiltonian1} to atomic coordinates
such that $\hbar=1=m_{\rm e}$ and
\begin{equation}\label{epsiDef}
\epsi =
\sqrt{\frac{m_{\rm e}}{m_{\rm n}}}\,,
\end{equation}
which
will be our small dimensionless expansion parameter. Then \eqref{Hamiltonian1} becomes
\begin{equation}\label{Hamiltonian2}
H^{\epsi} = - \textstyle\frac{1}{2} \epsi^2 \Delta_x + H_{\rm e}
(x)
\end{equation}
with the electronic Hamiltonian
\begin{equation}\label{elecHam}
H_{\rm e} (x) = -\textstyle\frac{1}{2} \, \Delta_y + V (x, y)\,,
\end{equation}
which depends through $V$ parametrically on $x$. For complex molecules, one is often forced to model only
a carefully chosen subset of degrees of freedom. We assume here that the final form is still as in
\eqref{Hamiltonian2} at the expense of a suitable modification of $H_{\rm e} (x)$. Implicitly with
\eqref{Hamiltonian2} it is required that the initial  wave function has a kinetic (and hence also a total)
energy which is bounded independently of $\epsi$. Thus the nuclei move slowly.

The electronic structure problem is the eigenvalue equation
\begin{equation}\label{EigenProb}
H_{\rm e} (x) \chi_j (x) = E_j (x) \chi_j (x)
\end{equation}
 with $\chi_j \in \H_{\rm f}$, the Hilbert space for the electronic degrees of freedom.
Here ``f'' is supposed to remind of fast. Since electrons are fermions, in our case $\H_{\rm f} = S_{\rm a} L^2 (
\R^{3 N} ) $ with $S_{\rm a}$ projecting onto the antisymmetric wave
functions. The eigenvectors in  \eqref{EigenProb} are normalized as
$\langle \chi_j (x) \, , \, \chi_{j'} (x) \rangle_{ \H_{\rm f}}=
\delta_{jj'}$ with respect to the scalar product in  $\H_{\rm f}$. Note that the eigenvectors are determined only up to a phase $\vartheta_j(x)$. Their smooth dependence on $x$ will be addressed with more care below. Generically, in addition to the bound states, $H_{\rm e}(x)$
has continuous spectrum.

We label the eigenvalues in \eqref{EigenProb} as
\begin{equation}\label{EigenLabels}
E_1 (x) \leq E_2 (x) \leq \ldots\,,
\end{equation}
including multiplicity. The graph of $E_j$ is the $j$-th Born-Oppenheimer energy surface.
As a rule they have a complex structure with many crossings and avoided crossings.
Let $\psi(x)$ be a nucleonic wave
function, $\psi \in \H_{\rm s}$, with ``s'' reminding of slow. For simplicity we take  $\H_{\rm s} = L^2 (\R^{3 K}) $, remembering that to impose the physically correct statistics for the nuclei requires extra considerations \cite{MeTr}. States with the
property that the electrons are  precisely in the $j$-th
eigenstate are then of the form
\begin{equation} \label{BandForm}
\psi (x) \, \chi_j (x, y) \,,
\end{equation}
which we can think of either as a wave function in the total
Hilbert space $\H = \H_{\rm s} \otimes \H_{\rm f}$ or as a wave function
for the electrons depending parametrically on $x$. In the latter
case we abbreviate as $\psi (x) \chi_j (x) \in \H_{\rm f}$ for each $x$.
With this notation the projection operator onto the states of the
form \eqref{BandForm} is given by
\begin{equation} \label{pro}
P_j = | \chi_j (x)\rangle\langle \chi_j (x) |\,.
\end{equation}
Since the $ \chi_j (x)$'s are orthonormal,  $P_j$ is indeed an orthogonal projection.

In a molecular collision or in excitations through a laser pulse
only a few energy surfaces take part in the subsequent
dynamics. Thus we take a set $I$ of adjacent Born-Oppenheimer
surfaces and call
\begin{equation}\label{BandProj}
P = \sum_{j \in I} \, | \chi_j (x)\rangle\langle \chi_j (x) |
\end{equation}
 the projection onto the relevant subspace
(or subspace of physical interest). To ensure that other bands are not
involved, we assume them to have a spectral gap of size $a_{\rm gap}>0$ away from the
energy surfaces in $I$, i.e.
\begin{equation}\label{Gap}
| E_i (x) - E_j (x) | \geq a_{\mathrm{gap}} \quad \mathrm{for\,\,all} \,\,
j \in I \, , \, i \in I^c\,.
\end{equation}
Also the
continuous spectrum is assumed to be at least $a_{\rm gap}$ away
from the relevant energy surfaces.

If $\epsi = 0$, then
\begin{equation}\label{EpsZero}
\E^{- \I H^0 t} \psi (x) | \chi_j (x) \rangle = \left( \E^{-\I E_j (x) t} \psi
(x) \right)\,| \chi_j (x) \rangle\,.
\end{equation}
However, the Laplacian $-\epsi^2 \Delta_x$ weakly couples  the $x$-fibers and $P \H$ is a
subspace which is {\em not} invariant under the true unitary
propagator exp$[ - \I H^{\epsi} t ]$. Still, we take
\begin{equation}\label{ProjHamil}
P H^{\epsi} P
\end{equation}
 as an
approximate Hamiltonian for the time evolution in the relevant subspace $P \H$.
To compute $P H^{\epsi} P$ we choose $\varphi, \psi \in P \H$, to say
$\psi(x,y) = \sum_{j \in I} \psi_{j}(x) \chi_j(x,y)$ with  $\psi_{j} \in  \H_{\rm s}$ and similarly for $\varphi$, and sandwich to the right and left
as
\begin{eqnarray}\label{PHP}
\langle\varphi, H \psi\rangle  &= &\sum_{m, n \in I} \, \int \D x\, \varphi_m (x)^\ast\langle \chi_m (x)
, \, H^{\epsi} \chi_n (x) \rangle_{\H_{\rm f}} \, \psi_n (x) \\
& = & \sum_{m, n \in I} \int \D x\, \varphi_m (x)^\ast \Big[ \delta_{m n} E_m
(x) + {\textstyle\frac{1}{2}} \epsi^2 \big( - \delta_{mn} \Delta - \nabla
\cdot \langle \chi_m (x), \nabla \chi_n (x) \rangle_{\H_{\rm f}} \nonumber \\
& &+ \, \langle \nabla \chi_m (x)  , \, \chi_n (x) \rangle_{\H_{\rm f}}\cdot \nabla + \langle
\nabla \chi_m (x) , \, \cdot\nabla \chi_n (x) \rangle_{\H_{\rm f}} \big) \Big] \psi_n (x)\,,\nonumber\\
&=& \int dx \phi_m(x)^\ast (P H^{\epsi} P)_{mn} \psi_n (x)\,,
\end{eqnarray}
where all derivatives are with respect to $x$. It is thus natural to introduce the geometric  phase,  or Berry connection,
\begin{equation}\label{BerryPhase}
\A_{m n} (x) =\I \langle \chi_m (x)\, , \, \nabla \chi_n (x)
\rangle_{\H_{\rm f}},\quad\A_{m n} (x)^\ast = \A_{n m} (x)\,.
\end{equation}
Here, for each $x$, $\chi_j(x), j\in I,$ is completed to an orthonormal basis through $\chi_j, j\in I^c$. We also define the nuclear momentum operator $p = - \I \epsi \nabla$. Noting
that $\langle \nabla \chi_m  , \, \chi_n\rangle  +\langle \chi_m   , \, \nabla
\chi_n \rangle = 0$ and inserting  in the last term of \eqref{PHP}, one
obtains
\begin{eqnarray}\label{PHP2}
(P H^{\epsi} P)_{mn} & = & E_m (x) \delta_{m n} + {\textstyle\frac{1}{2}}
\sum_{\ell \in I} \big( p\,\delta_{m \ell} -  \epsi \A_{m \ell} (x)\big)\cdot \big(
p \,\delta_{\ell n} -  \epsi \A_{\ell n} (x)\big)\nonumber\\
&  & +\,{\textstyle\frac{1}{2}} \epsi^2 \sum_{\ell \in I^c} \, \A_{m \ell}
(x)\cdot \A_{\ell n} (x) \,.
\end{eqnarray}
$P H^{\epsi} P$ acts on wave functions of the form $\psi_n (x)$, $n \in I$,  i.e.\
$\psi \in \H_{\rm s} \otimes \C^{| I|}$.

To make further progress we concentrate on two cases of physical interest.\\
1) Let $|I| = 1$, $I= \{j\}$, i.e., we consider a single nondegenerate energy band, which by assumption is isolated from the remaining energy bands. Then
$\chi_j(x)$ can chosen to be real and smooth, which implies
\begin{equation}\label{Berry}
 \A_{jj} (x) = 0\,.
\end{equation}
In other words, by a suitable choice of the phase $\vartheta_j(x)$, see below equation (\ref{EigenProb}), i.e., by a suitable {\it gauge},
the geometric phase can be made to vanish.
According to \eqref{PHP} this results in
\begin{equation}\label{ProjHamil1}
(P H^{\epsi} P)_{jj} = \textstyle\frac{1}{2}p^2 + E_j(x) + \textstyle\frac{1}{2} \epsi^2\phi(x)
\end{equation}
with the Born-Huang potential
\begin{equation}\label{BH}
\phi(x) = \langle \nabla\chi_j(x)\,,\,(1- |\chi_j(x)\rangle\langle\chi_j(x)|)\cdot\nabla\chi_j(x)\rangle_{\H_{\rm f}}\,.
\end{equation}
2) Let  $|I| = 2$. For convenience we label as $m,n=0,1$. $E_0(x)$ and $E_1(x)$
are allowed to cross. To be very specific, and to link to Section \ref{S3}, let us assume that $x \in
\R^{2}$ and $E_0(0) = E_1(0)$, while $E_0(x) \neq E_1(x)$ otherwise.
As first pointed out by Mead and Truhlar \cite{MeTr}, if one insists on smoothness of the eigenbasis away from $x=0$, then $\chi_0(x),
\chi_1(x)$ are necessarily complex-valued.
$ \A_{m n} (x)$, $m,n=0,1$,  corresponds to a vector potential with a magnetic field which is concentrated at $x=0$ and which
cannot be gauged away. As to be explained in more detail in Section \ref{S3}, in principle, there are then two choices. In the subspace $P\mathcal{H}$ one can work in the {\it adiabatic} basis, adopted here, at the expense of having singular coefficients at $x=0$. The alternative choice is to pick some {\it diabatic} basis, which is not an eigenbasis of $H_{\mathrm{e}}$, but has the advantage that  in this representation $P H^{\epsi} P$ depends smoothly on $x$.\medskip

The conventional derivation of the time-dependent Born-Oppenheimer approximation
leaves two points
in the dark.\medskip\\
(i) Since the total energy is fixed, the nuclei move slowly and thus one has to
follow their dynamics over a sufficiently long time to see a nontrivial dynamics. In
our units the velocities are of order $\epsi$, hence times have to be of order
$\epsi^{-1}$. At the zeroth order Born-Oppenheimer approximation one neglects  terms
of order $\epsi$ in the intra-band Hamiltonian $PH^\epsi P$, which over times of
order $\epsi^{-1}$ add up to an error of order 1 in the wave function. Thus on this
time-scale the   first-order Born-Oppenheimer approximation, i.e.\  the inclusion of
the Berry connection term, becomes mandatory. But there is a more subtle problem
with the standard derivation.  Since the commutator $[P,H^\epsi]$ is of order
$\epsi$, it is not clear a priori, cf.\ \eqref{Duhamel}, if
states of the form \eqref{BandForm} remain even approximately of this form for the relevant
times of order $\epsi^{-1}$,  put differently, if $P\H$ is approximately invariant under
the time evolution generated by $H^\epsi$ for sufficiently long times.
Again, this is because  naively the order $\epsi$  error in the generator may add up to an
 error of order $1$ during  times  of order $\epsi^{-1}$. Thus the substitution of $H^\epsi$ by $PH^\epsi P$
 needs justification.\medskip\\

(ii) If one wants to have the motion of nuclei at higher precision, then the simple
ansatz \eqref{BandForm} becomes questionable. Rather one expects small corrections
to the product form. From the way we have presented the computation it is not so
clear how to include such effects.\medskip

In Section~\ref{S2} we review mathematical results  which settle the two issues
raised. The justification of the first-order Born-Oppenheimer approximation as in
\cite{SpTe} is based on generalizing the standard adiabatic theorem of quantum
mechanics to a space-adiabatic theorem. The higher order corrections require a
systematic scheme developed in \cite{BrNo, EmWe, NeSo, MaSo, PST1, So} heavily based
on pseudo-differential calculus with operator valued symbols, a method which traces
back to the pioneering work of Sj\"{o}strand \cite{Sj}. We therefore refrain from
explaining the general scheme in detail, but instead present the main ideas and a
new elementary derivation of the correct second order Born-Oppenheimer Hamiltonian.
Section~\ref{S3} deals with applications where the higher order corrections are
potentially of importance. In particular we discuss the motion near a conical
intersection and the reactive scattering $H_2 + H \rightarrow H + H_2$.


\section{Justification of the Born-Oppenheimer approximation and higher order corrections}\label{S2}

In this section we explain several mathematical results concerning the validity of the Born-Oppenheimer approximation and higher order corrections to it. We do not restate precise mathematical theorems, which can be found in the quoted literature, but
focus on structural aspects instead. To this end we first argue how the gap in the conventional derivation of the Born-Oppenheimer approximation, as discussed  in the introduction,  can be closed. This understanding will open up the way to systematically determine also the higher order corrections to the Born-Oppenheimer approximation. While the general theory describing higher order corrections is rather technical,  in Section~\ref{SS22} we present an elementary computation which yields the second order in $\epsi$ corrections to the Born-Oppenheimer approximation.

\subsection{Justification of the time-dependent  Born-Oppenheimer approximation}\label{SS21}

In the conventional derivation of the zeroth and first order Born-Oppenheimer approximation the effective Born-Oppenheimer Hamiltonian is obtained by expanding $PH^\epsi P$ in powers of $\epsi$, where $P$ projects on wave functions of the form \eqref{BandForm}. As explained  under point (i) above, there is left open an important point in this derivation.
For $PH^\epsi P$ to be a valid effective Hamiltonian in the first place,  the dynamics generated by $PH^\epsi P$ for initial states in $P\H$ must be close to the one generated by the true Hamiltonian $H^\epsi$ for sufficiently long times, i.e.\ the difference
\[
\left(\E^{-\I H^\epsi t/\epsi} -\E^{-\I P H^\epsi P t/\epsi} \right)\,P
\]
must be small in an appropriate sense. The Duhamel formula yields
\begin{eqnarray} \label{Duhamel}
\left(\E^{-\I H^\epsi t/\epsi} -\E^{-\I P H^\epsi P t/\epsi} \right)\,P &=& \I \E^{-\I H^\epsi t/\epsi} \int_0^{t/\epsi}\D s\,  \E^{\I H^\epsi s} \left(  PH^\epsi P- H^\epsi  \right) \E^{-\I P H^\epsi P s}\,P  \nonumber
\\&=&
\I \E^{-\I H^\epsi t/\epsi} \int_0^{t/\epsi}\D s\,  \E^{\I H^\epsi s} \left(  PH^\epsi P- H^\epsi  \right) P\,\E^{-\I P H^\epsi P s} \nonumber
\\&=&
\I \E^{-\I H^\epsi t/\epsi} \int_0^{t/\epsi} \D s\,  \E^{\I H^\epsi s}\, [P,H^\epsi] \, P\,\E^{-\I P
H^\epsi P s}\,.
\end{eqnarray}
When acting on wave functions with kintic energy of order 1, i.e.\ $\|\epsi\nabla\psi^\epsi\|^2=\Or(1)$,  the commutator
\[
[P,H^\epsi] P= \left[ \sum_{j \in I} \, | \chi_j (x)\rangle\langle \chi_j (x) | , -{  \frac{\epsi^2}{2}}\Delta_x \right] P = \Or(\epsi)
\]
 contains terms of order $\epsi$ but not smaller,  independently of the choice of $\chi_j(x)$.
 Thus in \eqref{Duhamel} a naive estimate of the right hand side yields an error of order $1$. This is because the smallness of the integrand is cancelled by the growing domain of integration.
Hence it is not clear from a naive perturbation argument  that the dynamics generated $PH^\epsi P$ and thus the Born-Oppenheimer approximation is close to the true dynamics for sufficiently long times. This is directly linked to the question, whether the subspace $P\H$ of wave functions of the form \eqref{BandForm} is invariant under the true dynamics for sufficiently long times.
We stress this point, because the question of invariance of subspaces on which one approximates the dynamics by an effective Hamiltonian is crucial for understanding the higher order corrections to the Born-Oppenheimer approximation.

For the case of the first order Born-Oppenheimer approximation  it was shown in \cite{SpTe} how to use arguments similar to those used in the proof of the adiabatic theorem by Kato \cite{Ka} in order to prove approximate invariance of $P\H$. In such a {\it space}-adiabatic theorem, the positions $x$ of the nuclei take the role played by time $t$ in usual {\it time}-adiabatic theory. The mechanism is that the integrand in \eqref{Duhamel} is oscillatory and therefore the errors of order $\epsi$ in the integrand cannot add up to an error of order $1$ even for a domain of integration of size $1/\epsi$. The statement proved in \cite{SpTe} (with some typos in the proof corrected in \cite{Te}) is the following:
Let $P_E:={\bf 1}_{(-\infty,E]}(H^\epsi)$ be the spectral projection of $H^\epsi$ on energies below $E$. Then there exists a constant $C<\infty$ independent of $\epsi$ such that
\begin{equation}\label{St1}
\left\|\left(\E^{-\I H^\epsi t/\epsi} -\E^{-\I P H^\epsi P t/\epsi} \right)P\,P_E\right\|_{\B(\H)} \leq C\,\epsi\,(1+|t|)\,(1+|E|)\,,
\end{equation}
where $\|\cdot\|_{\B(\H)}$ is the norm of bounded operators.

Thus on subspaces of finite total energy the adiabatic approximation holds with a uniform error of order $\epsi$. Note that
 \eqref{St1} holds also with $P_E$ replaced by ${\bf 1}_{(-\infty,E]}(-\frac{\epsi^2}{2}\Delta)$,
 i.e.\ for finite kinetic energies of the nuclei. For large kinetic energies the velocities of the nuclei are no longer small compared to those of the electrons  and thus the adiabatic approximation breaks down.

Clearly the first order Born-Oppenheimer approximation now follows from \eqref{St1} by expanding the adiabatic Hamiltonian $PH^\epsi P$ as in \eqref{PHP}. If we define the first order Born-Oppenheimer Hamiltonian as
\begin{equation}\label{1stBO}
H_\epsi^{(1)} = \sum_{m,n\in I} |\chi_m(x)\rangle \Big(E_m (x) \delta_{m n} + {\textstyle\frac{1}{2}}
\sum_{\ell \in I} \big( p\,\delta_{m \ell} - \epsi \A_{m \ell} (x)\big)\cdot \big(
p \,\delta_{\ell n} -  \epsi \A_{\ell n} (x)\big)\Big)\langle\chi_n(x)|\,,
\end{equation}
where we omit one term of order $\epsi^2$ in \eqref{PHP}, then
\begin{equation}\label{St2}
\left\|\left(\E^{-\I H^\epsi t/\epsi} -\E^{-\I H_\epsi^{(1)} t/\epsi} \right)P\,P_E\right\|_{\B(\H)} \leq \tilde C\,\epsi\,(1+|t|)\,(1+|E|)\,.
\end{equation}
By now it is obvious why we cannot improve the approximation simply by adding the $\epsi^2$ terms in the expansion of $PH^\epsi P$ to the effective Hamiltonian. Their contribution is of the same order as the error in the adiabatic approximation  \eqref{St1}.  One might hope that the order of the error in \eqref{St1} can be improved by a more careful analysis. This is however not the case, as can be seen directly  from the proof in \cite{SpTe} or, alternatively, from the scheme to be presented in the following.  In particular there is no reason to expect that \eqref{PHP} is the correct second order Born-Oppenheimer Hamiltonian and indeed, it isn't.


\subsection{Corrections to the time-dependent Born-Oppenheimer approximation}\label{SS22}

But then the question arises whether and in which sense is the first order time-dependent Born-Oppenheimer approximation the leading term in a systematic perturbation expansion? This problem was considered several times in the literature and the solutions differ not only with respect to the level of mathematical rigor but also the formulation of the result itself is not unique. In the following we explain the approach  from \cite{PST1} which heavily relies on methods developed by Martinez, Nenciu, and Sordoni \cite{NeSo,MaSo}.
The main idea is to replace the subspace $P\H$ by a so called almost invariant subspace $P^\epsi\H$, which is invariant under the full dynamics to higher accuracy  than $P\H$. Because of the very technical character of the general construction, we only sketch the main ideas and instead give an alternative derivation of the correct second order Born-Oppenheimer Hamiltonian not using the pseudo-differential operator machinery.  We shortly comment on other approaches at the end of this section.

 Since  the range of $P$ is invariant only up to an error of order $\epsi$, the form \eqref{BandForm} is not adequate for higher order approximations. Instead one has to look for slightly tilted subspaces $P^\epsi\H$ which are invariant under the dynamics generated by $H^\epsi$ with smaller errors. Indeed one can find an orthogonal projection $P^\epsi$ close to $P$ such that for any $n\in\N$ there is a constant $C_n<\infty$ independent of $\epsi$ and $t$ satisfying
\begin{equation}\label{St3}
\left\|\left(\E^{-\I H^\epsi t/\epsi} -\E^{-\I P^\epsi H^\epsi P^\epsi t/\epsi} \right)P^\epsi\,P_E\right\|_{\B(\H)} \leq C_n\,\epsi^n\,|t|\,.
\end{equation}
It turns out that the projected Hamiltonian $P^\epsi H^\epsi P^\epsi$ can be expanded in powers of $\epsi$ as
\[
P^\epsi H^\epsi P^\epsi  =\sum_{j=0}^n \epsi^j H_j +\Or(\epsi^{n+1})=: H_{(n)}^\epsi + \Or(\epsi^{n+1})\,,
\]
which together with \eqref{St3} implies
\begin{equation}\label{St4}
\left\|\left(\E^{-\I H^\epsi t/\epsi} -\E^{-\I H_{(n)}^\epsi t/\epsi} \right)P^\epsi\,P_E\right\|_{\B(\H)} \leq \tilde C_n\,\epsi^n\,|t|\,.
\end{equation}
In the same fashion the projector $P^\epsi$ can be expanded in a Taylor series, which for the purpose of the estimates in (\ref{St3}) and (\ref{St4}) may be truncated at the appropriate order. Since  the zeroth and first order Born-Oppenheimer Hamiltonians are just the leading terms in the expansion of $H_{(n)}^\epsi$, one could call   $H_{(n)}^\epsi$ the $n$-th order Born-Oppenheimer Hamiltonian.
But  $H_{(n)}^\epsi$ is an effective Hamiltonian which acts on a $\epsi$-dependent subspace
of the full Hilbert space $\H$, while the usefulness of the standard Born-Oppenheimer approximation comes partly from the fact that the effective Hamiltonian acts on wave functions depending only on nuclear and possibly a few discrete electronic degrees of freedom. Before we explain how to remedy this shortcoming let us briefly explain how to construct the projector $P^\epsi$.

The construction of the projection $P^\epsi$ follows a general scheme outlined by
Emmerich and Weinstein \cite{EmWe}, refined in \cite{BrNo} and \cite{NeSo}, and
finally applied to the Born-Oppenheimer approximation by Martinez and Sordoni
\cite{MaSo}. The basic idea is to determine the coefficients $P_j$ in the expansion
$P^\epsi_{(n)}=P + \sum_{j=1}^n \epsi^j P_j$ order by order such that
$P^\epsi_{(n)}$ is approximately a projection and commutes with $H^\epsi$ up to
terms of order $\epsi^{n+1}$.
 Writing $P_0=P$ it clearly holds that $P_0$ is a projection and that
the commutator
\[
\left[ P_0, H^\epsi \right] = \left[ P_0, -{\frac{\epsi^2}{2}}\Delta \right] = \I \epsi (\nabla P_0) \cdot (-\I\epsi\nabla) + \frac{\epsi^2}{2}(\Delta P_0)
\]
is of order $\epsi$ when acting on functions of bounded kinetic energy.  One now determines $P_j$ inductively by requiring that
\begin{eqnarray}\label{Cond}
(P^\epsi_{(n)})^2 - P^\epsi_{(n)}& =& \Or(\epsi^{n+1})\,,\nonumber\\
\\
\left[ P^\epsi_{(n)},\,H^\epsi \right] &=& \Or(\epsi^{n+1})\,,\nonumber\,
\end{eqnarray}
and assuming the the analogous statement holds already for $P_{(n-1)}^\epsi$.
While the general construction is most conveniently done using the theory of $\epsi$-pseudodifferential operators, let us explicitly determine $P_1$ from the above conditions.
Because of
\[
(P_0+\epsi P_1)^2 - (P_0+\epsi P_1) = \epsi ( P_0P_1 + P_1P_0 -P_1) + \Or(\epsi^2)
\]
we must require that
\begin{equation}\label{ProjReq}
P_1 = P_0P_1 + P_1P_0 + \Or(\epsi)\,
\end{equation}
in order to make the order $\epsi$ term vanish.
And the order $\epsi$ term in
\[
\left[ P_0+\epsi P_1, H^\epsi \right] =  \I \epsi (\nabla P_0) \cdot (-\I\epsi\nabla)  +\epsi [P_1, H_{\rm e}] + \Or(\epsi^2)
\]
vanishes, if $P_1$ satisfies
\begin{equation}\label{P1Eq}
[P_1,H_{\rm e}] = \I(\nabla P_0)\cdot (\I\epsi\nabla) +\Or(\epsi)\,.
\end{equation}

To focus on the simplest case, from now on   we assume that $P_0(x)$ projects onto the eigenspace of a single eigenvalue $E_j(x)$, i.e.\ that we are in the situation of item 1) in the introduction.
In this special but most important case, the unique solution of \eqref{P1Eq} up to order $\epsi$ is obtained by multiplying \eqref{P1Eq} from the left and from the right by $P_0$ resp.\ $(1-P_0)$. The block-diagonal terms are of order $\epsi$, since $P_0( \nabla P_0 )P_0 = (1-P_0)(\nabla P_0)(1-P_0)=0$. For the off-diagonal terms one can invert  $(H_{\rm e}-E_j)$ on the range of $(1-P_0)$ and finds that
\[
P_1 = \I P_0(\nabla P_0) (H_{\rm e}-E_j)^{-1}(1-P_0)\cdot (\I\epsi\nabla) + \mbox{adj.} =   \I(\I\epsi\nabla)\cdot P_0(\nabla P_0) (H_{\rm e}-E_j)^{-1}(1-P_0)+ \mbox{adj.}  +\Or(\epsi)\,,
\]
which also satisfies the requirement \eqref{ProjReq}. The abbreviation $+$ adj.\ means that the adjoint of everything to the left is added. Here and in the following we use the fact that the spectral projection $P_0(x)$ and the reduced resolvent $(H_{\rm e}(x) -E_j(x))^{-1}(1-P_0(x))$ are smooth function of $x$ with values in the bounded operators on the electronic Hilbert space $\H_{\rm f}$. This follows from the assumed smoothness of $H_{\rm e}(x)$ and the gap condition. As a consequence the commutator of the momentum operator $p=-\I\epsi\nabla$ with any such operator yields only a lower order term in $\epsi$, a fact which will be used several times in the following computations.

In a next step the operator $P_{(1)}^\epsi$ can be turned into a true orthogonal projection by adding a term of order $\epsi^2$, see \cite{NeSo}, which we denote again by $P_{(1)}^\epsi$.
Now the question arises,  whether the range of $P_{(1)}^\epsi$ is invariant under the dynamics generated by $H^\epsi$ up to errors of order $\epsi^2$. A naive perturbation argument as in \eqref{Duhamel} will only yield an error of order $\epsi$, which, however, could be improved using again the space-adiabatic approach of \cite{SpTe}. Alternatively the results of \cite{MaSo,PST1} show that
 instead of
\eqref{St1} we now have
\begin{equation}\label{P1diag}
\left\|\left(\E^{-\I H^\epsi t/\epsi} -\E^{-\I P^\epsi_{(1)} H^\epsi P^\epsi_{(1)} t/\epsi} \right)P^\epsi_{(1)}\,P_E\right\|_{\B(\H)} \leq C\,\epsi^{2}\,(1+|t|)(1 + |E|)\,.
\end{equation}
With \eqref{P1diag} we are in a position to determine an effective Hamiltonian on the range of $P_{(1)}^\epsi$ by expanding $P^\epsi_{(1)} H^\epsi P^\epsi_{(1)}$ in powers of $\epsi$ and keeping terms up to order $\epsi^2$. This improves the conventional first order Born-Oppenheimer approximation by one order.
However, because of the momentum operator appearing in $P_1$, the range of $P_{(1)}^\epsi$ is not spanned by wave functions of the form $\psi(x)\chi_j^\epsi(x,y)$ as in
\eqref{BandForm}. Therefore $ P^\epsi_{(1)} H^\epsi P^\epsi_{(1)} $ can no longer be seen as an effective Hamiltonian acting on a nucleonic wave function $\psi(x)$ alone. But this reduction in the degrees of freedom of the state space is the crucial feature which makes the first order Born-Oppenheimer approximation so useful. In order to retain this feature also for the higher order Born-Oppenheimer approximations we thus need to map the range of $P^\epsi_{(1)}$ unitarily to the space $L^2(\R^{3K})$ of nucleonic wave functions in the case of a simple electronic band or to $L^2(\R^{3K}, \C^{|I|})$ in the case of a group of $|I|$ bands.
In \cite{PST1} we construct a unitary operator $U^\epsi:P^\epsi\H\to L^2(\R^{3K}, \C^{|I|})$ and define the effective Born-Oppenheimer Hamiltonian for the group of levels in $I$ as
\begin{equation}\label{generalBO}
H^\epsi_{\rm BO} = U^\epsi\,P^\epsi\, H^\epsi\,P^\epsi\,U^{\epsi\,*}\,,
\end{equation}
i.e.\ the full Hamiltonian $H^\epsi$ is projected onto the almost invariant subspace associated with the  levels in $I$ and then unitarily mapped to the $\epsi$-independent space of nucleonic wave functions. Thus \eqref{St3} becomes
\begin{equation}\label{BOSt}
\left\|\left(\E^{-\I H^\epsi t/\epsi} -U^{\epsi\,*}\,\E^{-\I H^\epsi_{\rm BO} t/\epsi} U^\epsi \right)P^\epsi\,P_E\right\|_{\B(\H)} \leq C_n\,\epsi^n\,|t|\,,
\end{equation}
where  $\E^{-\I H^\epsi_{\rm BO} t/\epsi}$ is the effective Born-Oppenheimer propagator  of the nuclei within the relevant group of bands up to any order in $\epsi$. The expansion of $H_{\rm BO}^\epsi$ in powers of $\epsi$  yields at the  leading orders the usual zeroth and first order Born-Oppenheimer Hamiltonian, i.e.\ the matrix in the brackets in \eqref{1stBO}, but in addition also the correct higher order terms.

In order to avoid the technicalities of the general construction, let us determine the correct second-order Born-Oppenheimer Hamiltonian for the     special case of a singe simple level $E_j(x)$ by an elementary calculation.
For this case we already  computed $P^\epsi=P_0+\epsi P_1+\Or(\epsi^2)$   with $P_0(x) = |\chi_j(x)\rangle\langle \chi_j(x)|$ and
\[
  P_1 = -\I(-\I\epsi\nabla)\cdot P_0(\nabla P_0) (H_{\rm e}-E_j)^{-1}(1-P_0)+ \mbox{adj.}  =:p\cdot B+B^*\cdot p\,.
\]
Here and in the following we abbreviate $p= -\I\epsi\nabla_x$ for the momentum operator.
Also for $U^\epsi$ we make the ansatz $U^\epsi = U_0 + \epsi U_1 +\Or(\epsi^2)$ where clearly
\begin{equation}\label{U_0 for m=1}
U_0(x) = \langle \chi_j(x)|
\end{equation}
is unitary from $P_0\H$ to $L^2(\R^{3K})$. Note that at this point the construction of $U^\epsi$ is not unique since the choice of basis $\chi_j(x)$ enters.

Without loss of generality we write $U_1=U_0 A$ for some operator $A$ on $\H$. Then the requirement that $U^{\epsi\,*}$ is unitary up to $\Or(\epsi^2)$
yields
\begin{equation}\label{UCond}
(U_0+\epsi U_1)(U_0^* +\epsi U_1^*) = 1 +\epsi(U_0A^*U_0^* + U_0AU_0^*) +\Or(\epsi^2) \stackrel{!}{=}1+\Or(\epsi^2)\,.
\end{equation}
The requirement that the range of $U^{\epsi\,*}$ is the range of $P^\epsi$ up to $\Or(\epsi^2)$ yields
\[
(1- P_0-\epsi P_1)(U_0^* +\epsi U_1^*)= \epsi (A^*-P_0 A^*-P_1)U_0^* +\Or(\epsi^2)\stackrel{!}{=}\Or(\epsi^2)\,,
\]
i.e.\ that
\[
(1-P_0)A^* = P_1P_0+\Or(\epsi)\,.
\]
A  solution of this equation is
$A^*=B^*\cdot p$, which also makes the $\Or(\epsi)$ term in \eqref{UCond} vanish. Thus
\[
U_1 = U_0 \,\,p\cdot B
\]
yields an almost unitary $U^\epsi_{(1)}= U_0 + \epsi U_1$ that almost intertwines $P^\epsi_{(1)}\H$ and $L^2(\R^{3K})$. It is important for the following that
 $U^\epsi_{(1)}$ can be modified by a term $\epsi^2U_2^\epsi$ of order $\epsi^2$ (i.e.\ $U^\epsi_2$ is $\Or(1)$) to make it a true unitary exactly intertwining $P^\epsi_{(1)}\H$ and $L^2(\R^{3K})$. As in the case of $P^\epsi_{(1)}$ we denote the modified operator $U^\epsi_{(1)}= U_0 +\epsi U_1 +\epsi^2 U_2^\epsi$ again by the same symbol. It is shown in \cite{PST1} how to construct
 $U_2^\epsi$  explicitly, but its exact form is not important for the following.  We will only use that
 unitarity of the modified  $U_{(1)}^\epsi$ together with the fact that $U_1U^{*}_0 + U_0U_1^* =U_0 P_1 U_0^*=0$ implies that
 \begin{equation}\label{unitarity}
 U_1U_1^* + U_0U_2^{\epsi\,*} + U_2^\epsi U_0^* =\Or(\epsi)\,.
 \end{equation}
We thus have that
\begin{equation}\label{Appr1}
\E^{-\I P^\epsi_{(1)} H^\epsi P^\epsi_{(1)} t/\epsi} P_{(1)}^\epsi =    U^{\epsi\,*}_{(1)}\,U^\epsi_{(1)} \,\E^{-\I  P^\epsi_{(1)} H^\epsi P^\epsi_{(1)} t/\epsi}U^{\epsi\,*}_{(1)}\, U^\epsi_{(1)} P_{(1)}^\epsi
  =U^{\epsi\,*}_{(1)}\,\E^{-\I  U^\epsi_{(1)} P^\epsi_{(1)} H^\epsi P^\epsi_{(1)}U^{\epsi\,*}_{(1)} t/\epsi} U^\epsi_{(1)} P_{(1)}^\epsi
\end{equation}
and therefore   with \eqref{P1diag} that \eqref{Appr1} approximates  the true time evolution up to errors of order $\epsi^2$,
\begin{equation}\label{UPHPU}
\left\|\left(\E^{-\I  H^\epsi   t/\epsi}  - U^{\epsi\,*}_{(1)}\,\E^{-\I  U^\epsi_{(1)} P^\epsi_{(1)} H^\epsi P^\epsi_{(1)}U^{\epsi\,*}_{(1)} t/\epsi} U^\epsi_{(1)}  \right)P^\epsi_{(1)}\,P_E\right\|_{\B(\H)} \leq C\,\epsi^{2}\,(1+|t|)\,.
\end{equation}
Hence we can now expand $U^\epsi_{(1)} P_{(1)}^\epsi H^\epsi P_{(1)}^\epsi U^{\epsi\,*}_{(1)}$ in powers of $\epsi$ to obtain the second order Born-Oppenheimer Hamiltonian, which now acts on the $\epsi$-independent space $L^2(\R^{3K})$ of nucleonic wave functions. The expansion yields
\begin{eqnarray}
U^\epsi_{(1)} P_{(1)}^\epsi H^\epsi P_{(1)}^\epsi U^{\epsi\,*}_{(1)}&=&U^\epsi_{(1)}  H^\epsi   U^{\epsi\,*}_{(1)}=(U_0+\epsi U_1+\epsi^2 U_2^\epsi)\left( {\textstyle\frac{p^2}{2}} + H_{\rm e}\right) (U_0^*+\epsi U_1^*+\epsi^2 U_2^{\epsi\,*}) +\Or(\epsi^3)\nonumber\\
&=& U_0 \left( {\textstyle\frac{p^2}{2}}  + H_{\rm e}\right)  U_0^*\label{0Or}\\
&&+\,\, \epsi\left( U_0 \left({\textstyle\frac{p^2}{2}} + H_{\rm e}\right) U_1^* +\mbox{adj.}\right)\label{1Or}\\
&&+\,\,\epsi^2 U_1\left( {\textstyle\frac{p^2}{2}} + H_{\rm e}\right) U_1^*\label{2Or}\\
&&+\, \epsi^2\left( U_0\, \left({\textstyle\frac{p^2}{2}} + H_{\rm e}\right)\,U_2^{\epsi\,*} + U_2^\epsi\, \left({\textstyle\frac{p^2}{2}} + H_{\rm e}\right)\,U_0^*\right)+\Or(\epsi^3)\,.\label{2OrB}
\end{eqnarray}
We evaluate the four terms \eqref{0Or}--\eqref{2OrB} separately. Expanding \eqref{0Or} yields
\begin{eqnarray}
U_0 \left({\textstyle\frac{p^2}{2}} + H_{\rm e}\right)  U_0^*&=& {\textstyle\frac{p^2}{2}} + E_j + U_0\left[{\textstyle\frac{p^2}{2}} ,U_0^*\right]\nonumber\\
&=& {\textstyle\frac{p^2}{2}} + E_j -\I\epsi U_0(\nabla U_0^*)\cdot p  -{\textstyle\frac{\epsi^2}{2}}U_0(\Delta U_0^*) \nonumber\\
&=&{\textstyle\frac{p^2}{2}} + E_j -\epsi \mathcal{A}\cdot p - {\textstyle\frac{\epsi }{2}} [p,\mathcal{A}] + {\textstyle\frac{\epsi^2 }{2}}\langle\nabla\chi_j,\cdot\nabla\chi_j\rangle\nonumber\\
&=& {\textstyle\frac{1}{2}}(p-\epsi\mathcal{A})^2 + E_j  + {\textstyle\frac{\epsi^2 }{2}} \phi\,,
\end{eqnarray}
where we abbreviated
\begin{equation}\label{ADef}
\mathcal{A}(x) = \I\epsi \langle \chi_j(x),\nabla\chi_j(x)\rangle
\end{equation}
 for the Berry connection coefficient and
 \begin{equation}\label{phiDef}
 \phi(x) = \langle \nabla\chi_j(x),\cdot(1-P_0)\nabla\chi_j(x)\rangle
 \end{equation}
  for the Born-Huang potential. This is, as expected, exactly  the usual Born-Oppenheimer Hamiltonian \eqref{PHP}. However, as explained before, in order to obtain the correct second order Born-Oppenheimer Hamiltonian, we also must take into account the terms \eqref{1Or}--\eqref{2OrB} which stem from the fact, that not $P\H$ but $P_{(1)}^\epsi\H$ is the correct adiabatically  invariant subspace at that order.
Expanding \eqref{1Or} yields
\begin{eqnarray}
U_0 \left({\textstyle\frac{p^2}{2}} + H_{\rm e}\right) U_1^* + \mbox{adj.} &=&
U_0\left( {\textstyle\frac{p^2}{2}}\, B^*\cdot p + p\cdot B\, {\textstyle\frac{p^2}{2}} \right) U_0^*+ U_0\left(H_{\rm e} B^*\cdot p + p\cdot B\, H_{\rm e}\right)U_0^*\nonumber\\
&=& U_0\left( [{\textstyle\frac{p^2}{2}}, B^*] \cdot p + p\cdot [B, {\textstyle\frac{p^2}{2}}] \right) U_0^*\nonumber\\
&=& -\I\epsi U_0\left( p \cdot \nabla B^* \cdot p - p\cdot\nabla B\cdot p \right) U_0^* +\Or(\epsi^2)\nonumber\\
&=& -2 \epsi\,U_0 \,p\cdot P_0 (\nabla P_0) (H_{\rm e}-E_j)^{-1}  (\nabla P_0) P_0 \cdot p\, U_0^* +\Or(\epsi^2)\nonumber\\
&=:& -2\epsi \mathcal{M}(x,p)+\Or(\epsi^2)\,,
\end{eqnarray}
where we used that $\nabla B P_0= \I P_0 (\nabla P_0)(H_{\rm e}-E_j)^{-1}(\nabla P_0) P_0$ and $P_0 \nabla
B^* = - \I P_0 (\nabla P_0)(H_{\rm e}-E_j)^{-1}(\nabla P_0) P_0$. Recall that the differences in operator
ordering between $p=-\I\epsi\nabla_x$  and operators depending on $x$ are of lower order in $\epsi$.  So any other operator ordering for the
quantization of $\mathcal{M}(x,p)$ than the one used in \eqref{1Or} works as well at the given order. Put differently, any quantization rule for the symbol (=function)
\begin{equation}\label{MDef}
\mathcal{M}(x,p) = U_0 \,p\cdot P_0 (\nabla P_0) (H_{\rm e}-E_j)^{-1}  (\nabla P_0) P_0 \cdot p\, U_0^*= \left\langle\, \nabla\chi_j\cdot p,(H_{\rm e}-E_j)^{-1}(1-P_0)\,p\cdot\nabla\chi_j\right\rangle_{\H_{\rm f}}
\end{equation}
will do the job.
The
simplest symmetric choice for $\mathcal{M}$ is presumably
\[
( \mathcal{M}\psi)(x) = \sum_{\ell,k=1}^{3K} \frac{1}{2}\Big( \mathfrak{m}_{\ell k}(x) (-\I\epsi \partial_{x_\ell}) (-\I\epsi \partial_{x_k}) +  (-\I\epsi \partial_{x_\ell}) (-\I\epsi \partial_{x_k})  \mathfrak{m}_{\ell k}(x)\Big)\psi(x)\,,
\]
where $\mathfrak{m}$ is the $x$-dependent matrix
\[
\mathfrak{m}_{\ell k} (x) =  \left\langle\, \partial_\ell \chi_j(x),(H_{\rm e}(x)-E_j(x))^{-1}(1-P_0(x))\,\partial_k \chi_j(x)\right\rangle_{\H_{\rm f}}
\]
For \eqref{2Or} we find
\begin{eqnarray}\label{2OrComp}
U_1({\textstyle\frac{p^2}{2}} + H_{\rm e}) U_1^*&=& U_0 \,p\cdot B ({\textstyle\frac{p^2}{2}}+H_{\rm e}) B^*\cdot p  \,U_0^*\nonumber\\
&=&U_0 \, p\cdot B (H_{\rm e}-E_j) B^*\cdot p \, U_0^* + U_0 \, p\cdot B ({\textstyle\frac{p^2}{2}}+E_j) B^*\cdot p \, U_0^* \nonumber\\&=& U_0 \,p\cdot P_0 (\nabla P_0) (H_{\rm e}-E_j)^{-1}  (\nabla P_0) P_0 \cdot p\, U_0^* +U_0 \, p\cdot B B^*\cdot p \, U_0^*\, ({\textstyle\frac{p^2}{2}}+E_j) +\Or(\epsi)\nonumber
\\&=& \mathcal{M}(x,p)+U_1  U_1^*\, ({\textstyle\frac{p^2}{2}}+E_j) +\Or(\epsi)
\,.
\end{eqnarray}
Finally the term \eqref{2OrB} cancels the second remaining term in \eqref{2OrComp}:
\begin{eqnarray}\label{2OrCompB}
 U_0\, \left({\textstyle\frac{p^2}{2}} + H_{\rm e}\right)\,U_2^{\epsi\,*} + U_2^\epsi\, \left({\textstyle\frac{p^2}{2}} + H_{\rm e}\right)\,U_0^* &=&  U_0\, \left({\textstyle\frac{p^2}{2}} + E_j\right)\,U_2^{\epsi\,*} + U_2^\epsi\, \left({\textstyle\frac{p^2}{2}} + E_j\right)\,U_0^*\nonumber\\
 &=& \left( U_0\, U_2^{\epsi\,*} + U_2^\epsi\,U_0^*\right)
 \left({\textstyle\frac{p^2}{2}} + E_j\right)+\Or(\epsi) \nonumber\\
 &=& - U_1  U_1^*\, ({\textstyle\frac{p^2}{2}}+E_j) +\Or(\epsi)\,,
 \end{eqnarray}
where we used \eqref{unitarity}.
Collecting all the results we find for the second order Born-Oppenheimer Hamiltonian of a simple isolated eigenvalue band $E_j$ that
\begin{equation} \label{BO2}
H_{\rm BO}^\epsi = {\textstyle\frac{1}{2}}(p-\epsi\mathcal{A}(x))^2 + E_j(x)  + {\textstyle\frac{\epsi^2 }{2}} \phi(x) - \epsi^2 \mathcal{M}(x,p)\,,
\end{equation}
where we recall that $p=-\I\epsi\nabla_x$ and that $\mathcal{A}(x)$, $\phi(x)$ and $\mathcal{M}(x,p)$ are defined in \eqref{ADef}, \eqref{phiDef} and \eqref{MDef}.

The relation between the dynamics generated by the Born-Oppenheimer Hamiltonian $H^\epsi_{\rm BO}$ and the true time evolution is now
\begin{equation}\label{HBO}
\left\|\left(\E^{-\I  H^\epsi   t/\epsi}  - U^{\epsi\,*}_{(1)}\,\E^{-\I  H^\epsi_{\rm BO} t/\epsi} U^\epsi_{(1)}  \right)P^\epsi_{(1)}\,P_E\right\|_{\B(\H)} \leq C\,\epsi^{2}\,(1+|t|)\,.
\end{equation}
Put differently, one can construct approximate solutions $\Psi_{\rm BO}(t)$ to the full molecular Schr\"odinger equation
\begin{equation}\label{SE}
\I\epsi\partial_t \Psi(t,x,y) = H^\epsi \Psi(t,x,y)\,,\qquad \Psi(t) \in L^2(\R^{3K})\otimes \H_{\rm f}
\end{equation}
by solving the effective Born-Oppenheimer Schr\"odinger equation  for the nuclei only
\begin{equation}\label{SEBO}
\I\epsi\partial_t \psi(t)= H^\epsi_{\rm BO} \psi(t)\,,\qquad\psi(t)\in L^2(\R^{3K})\,,
\end{equation}
and defining $\Psi_{\rm BO}(t) = U^{\epsi\,*}_{(1)} \psi(t)$. The difference between $\Psi_{\rm BO}(t)$
and the true solution of \eqref{SEBO} with the same initial condition $\Psi(0) = \Psi_{\rm BO}(0)\in P^\epsi_{(1)} P_E \H$ is of
order $\epsi^2$ in the norm of $\H$. However, there are many cases where it suffices to know $\psi(t)$ and
it is not necessary to map it back to a full molecular wave function $\Psi_{\rm BO}(t)$. For example for
the position distribution of the nuclei one has that
\[
\langle \Psi_{\rm BO}(t,x),  \Psi_{\rm BO}(t,x) \rangle_{\H_{\rm f}} =\langle U^{\epsi\,*}_{(1)}\psi(t,x) ,    U^{\epsi\,*}_{(1)}\psi(t,x) \rangle_{\H_{\rm f}}= |\psi(t,x)|^2 + \Or(\epsi^2)\,.
\]

We continue with several remarks on the second order Born-Oppenheimer Hamiltonian. The correct second order Born-Oppenheimer Hamiltonian \eqref{BO2} contains the additional term $\epsi^2\mathcal{M}(x,p)$ as compared to the conventional expression containing terms of that order, see \eqref{ProjHamil1}, \eqref{BH}. In contrast to the Born-Huang potential  $\phi(x)$, $\mathcal{M}(x,p)$ introduces a velocity-dependent correction in the form of an $x$-dependent effective mass tensor  $\mathfrak{m}$. In Section~\ref{S3} we compute $H_{\rm BO}^\epsi$ explicitly near a conical intersection of electronic energy levels, where the corrections to the conventional first order Born-Oppenheimer approximation become important.
A further issue is the gauge invariance of $H_{\rm BO}^\epsi$.  The Berry connection coefficient $\mathcal{A}(x)$ clearly depends on the choice of $\chi_j(x)$ and, in our particular context, can be made to vanish by a suitable choice of phase. However, although defined through $\chi_j(x)$, the  terms of order $\epsi^2$ are gauge invariant, as can be seen from writing them only by means of the projection $P_0(x)$,
\begin{eqnarray}
\phi(x)&=& {\rm Tr}_{\H_{\rm f}}  \Big(  \nabla P_0(x)\cdot\nabla P_0(x) \,(1-P_0(x) \Big)\,,\\
\mathcal{M}(x,p)&=&  {\rm Tr}_{\H_{\rm f}} \Big(  \big(p\cdot\nabla P_0(x)  \big)^2   \big(H_{\rm e}(x)-E_j(x)  \big)^{-1}  \big(1-P_0(x)  \big) \Big)\,.
\end{eqnarray}

In general, the structure of $H_{\rm BO}^\epsi$ is that of a semiclassical Hamiltonian for the nuclei,
since the momentum operator still carries the small parameter $\epsi$. Therefore it is natural to study in
a second step the semiclassical limit of the nuclear Schr\"odinger equation \eqref{SEBO}. There is a
variety of methods available for such a semiclassical analysis, most prominently the WKB approximation,
semiclassical wave packets and Wigner functions. All these techniques have been applied to further study
the standard Born-Oppenheimer approximation  and can also be used to investigate the higher order
Born-Oppenheimer approximations. But since the semiclassical limit   is conceptually and mathematically
different from the adiabatic approximation considered here, we do not further comment on it.

We conclude this section with a few remarks on the literature. In the physics
literature the correct second order Born-Oppenheimer Hamiltonian \eqref{BO2} was
first obtained by  Weigert and Littlejohn  in \cite{WeLi} for the case of matrix
valued $H_{\rm e}(x)$. They approximately diagonalize  the Hamiltonian $H^\epsi$ on
the full space $\H$, while we first reduce to an appropriate adiabatic subspace
corresponding to the electronic levels of interest and then approximate the
Hamiltonian on that subspace. This has the advantage that we can allow for $H_{\rm
e}(x)$ having continuous spectrum or level crossings outside the spectral part of
interest.

Another mathematical approach to the time-dependent Born-Oppenheimer approximation,
historically the first one, is due to Hagedorn and later Hagedorn and Joye
\cite{Ha1,Ha2a, Ha2b, HaJo}. In their approach the goal is not to construct an
approximate Born-Oppenheimer Hamiltonian but to find directly approximate solutions
of the molecular Schr\"odinger equation. This is achieved by approximating the
nucleonic wave function by localized semiclassical wave packets. Hence, in the
Hagedorn/Joye approach the adiabatic and the semiclassical approximation are done in
one package.\bigskip







\section{Dynamics near conical crossings}\label{S3}

The aim of this section is twofold. In the first part we show how the previous
scheme extends to a family of energy bands, yielding  a multiband effective
Hamiltonian. While this effective Hamiltonian is the starting point for an analysis
of the propagation of the wavefunction through  eigenvalue crossings, we will not
address this interesting problem here and instead refer the reader to the literature
\cite{FeGe, FeLa, Ha3, LaTe}, see also \cite{CLP} for a related result in a
time-independent setup.

In the second part, Section~\ref{Sec crossings}, we return to the one-band
Born-Oppenheimer dynamics and study in a simple case the behavior of the Berry
connection $\A$ and of the second-order corrections $\phi$ and $\M$ near a conical
crossing. We argue that this behavior is, in a sense, universal and leads
potentially to observable effects in chemical exchange reactions.

\subsection{The multiband effective Hamiltonian}  \label{Sec reduction}

We consider a family of $\m$ eigenvalue bands $\{ E_j \}_{j \in I}$,  $|\,I \,| = \m$, which may cross
each other but which are separated by a gap from the rest of the spectrum, see (\ref{Gap}). We denote as
$P_0(x)$ the eigenprojector of $H_{\rm e}(x)$ corresponding to such a family of bands, with $\dim \Ran
P_0(x) = \m$.

\medskip

By the construction outlined in the previous section, to such a family of bands there corresponds an
almost-invariant subspace. More precisely one constructs an orthogonal projector $P^{\e} = P_0 + \O(\e)$
satisfying \eqref{St3}. In order to describe  the dynamics inside the almost-invariant subspace in a
physically transparent way, one constructs a unitary operator $U^{\e}$ which intertwines
 $\Ran P^{\e}$ and a reference space, \ie
\begin{equation} \label{U intertwines}
U^{\e}: \Ran P^\e \to \H_{\rm ref}:= L^2(\R^{3K}, \C^\m).
\end{equation}
To the lowest order  $U^{\e} = U_0 + \O(\e)$, with
\begin{equation}\label{U0 definition}
    U_0(x) = \sum_{a = 1}^{\m} \left| e_a \> \<\ph_a(x)\right|,
\end{equation}
where $\{ e_a \}_{a \indexm}$ is the canonical basis in $\C^m$ and $\{ \ph_a(x)\}_{a \indexm}$ is any
orthonormal basis spanning $\Ran P_0(x)$ and depending smoothly on $x$. The freedom in choosing such a
basis is an additional feature with respect to the previous section. If $\{\ph_a \}_{a \indexm}$ is any
such basis, by the construction in \cite{PST1} one obtains a self-adjoint operator $H^{\e}_{\rm BO}$, such
that (\ref{BOSt}) is satisfied. It has the  asymptotic expansion
\begin{equation}\label{Effective m>1}
    \begin{array}{ccl}
      H^{\e}_{\rm BO} & := & U^\e \, P^\e \, H^\e \, P^\e \, U^{\e \, *} \\
                  &  = & \frac{1}{2}p^2 + W(q) +
                   \frac{\e}{2} \( p \cdot \A(x) + \A(x) \cdot p \) + \O(\e^2),
    \end{array}
\end{equation}
where we introduced the $\m \times \m$ matrices
\begin{equation}\label{W and A matrixes}
   \begin{array}{cclr}
     W(x)_{ab} &=& \< \ph_a(x), \, H_{\rm e}(x) \, \ph_b(x) \>_{\H_{\rm f}},& \\
                &&                                                      &  \\
     \A(x)_{ab}&=& i \< \ph_a(x), \nabla \ph_b(x) \>_{\H_{\rm f}},       &  a,b \indexm.
   \end{array}
\end{equation}

\noindent The eigenvalues of $W(x)$ are the energies $\{ E_j(x) \}_{j \in I}$ of the electronic
Hamiltonian $H_{\rm e}(x)$. The expansion in the second line of (\ref{Effective m>1}) is the
generalization of (\ref{1stBO}) to the case of an arbitrary basis.

Since $H_{\rm e}$, as defined in (\ref{elecHam}), satisfies time-reversal symmetry the eigenfunctions
$\{\ph_a \}_{a \indexm}$ can be chosen to be real valued. With this restriction $W(x)$ becomes a real
symmetric matrix and $ i \A(x)$ becomes a real antisymmetric matrix, in particular with vanishing diagonal
elements.

\medskip

The effective Hamiltonian (\ref{Effective m>1}) is clearly not unique. Indeed, one can consider a
different basis $\{\tilde{\ph}_a \}_{a \indexm}$, with
\[
\ph_a(x) = \sum_{b=1}^{\m} \, G(x)_{ab} \, \widetilde{\ph}_b(x)
\]
where $G(x)$ is some orthogonal matrix depending smoothly on $x$. The elements of the effective
Hamiltonian in these two basis are related by
\begin{eqnarray}
  \widetilde{W}(x) &=& G(x)^{-1} \, W(x) \, G(x),    \label{W tranforms}\\
  \widetilde{\A}(x) &=& G(x)^{-1} \, \A(x)\, G(x) + G(x)^{-1} \, \nabla G(x).  \label{A transforms}
\end{eqnarray}

\medskip

Notice that, even with time-reversal symmetry, in general the Berry connection $\A$ cannot be removed by a
change of  basis (a \emph{non-abelian} change of gauge). Indeed, by (\ref{A transforms}) this is possible
if and only if there exist a smooth solution $G(x)$ of the system of differential equations
\[
\d_i G(x) = - \A_i(x)  \, G(x), \qquad  i=1, \ldots, 3K,
\]
which as a necessary condition requires the vanishing of the corresponding matrix valued field
\[
\omega_{ij} = -\I \( \d_i \A_j - \d_j \A_i \) + \A_j \A_i - \A_i \A_j.
\]

\bigskip

\noindent The non-uniqueness of the effective multiband Hamiltonian (\ref{Effective m>1}) is irrelevant as
far as expectation values of physical observables are concerned, since a change of basis affects both the
Hamiltonian operator and the operator representing the observable quantity. However, the dependence on the
basis may be relevant if one considers $H^{\e}_{\rm BO}$ as a {\it bona fide} model Hamiltonian to
investigate the dynamics of the wavefunction in a family of bands. We will comment on this point in the
following.


\subsection{Higher-order corrections near conical crossings}
\label{Sec crossings}

We come back to the \emph{one-band} Born-Oppenheimer dynamics, \ie \, $ I = \{j\}$,
but consider now the case when the relevant energy band $E_j$ crosses other bands,
so that (\ref{Gap}) does not hold. The dynamics of a wavepacket localized far away
from the crossing points, and in the almost-invariant subspace corresponding to the
relevant energy band $E_j$, is approximately governed by the Born-Oppenheimer
Hamiltonian (\ref{BO2}). (While in Section~\ref{S2} we discussed only the case of a
globally isolated energy band, the more general case has been studied in \cite{PST1,
Te}). Since the adiabatic approximation breaks down near conical crossings one
expects that the higher order corrections $\A$, $\phi$ and $\mathcal{M}$ diverge
near such crossings. It has been pointed out by Berry and Lim \cite{BeLi} that the
Born-Huang potential $\phi$ creates a diverging repulsive force at the crossing
points, and some authors argued that the wavefunction must vanish at those points.
While it is certainly true that $\phi$ leads to a  repulsive force, we emphasize
that a complete analysis should take into account all contributions coming from the
second order corrections.

\medskip

Since a crossing involves generically just two bands, we focus on a family of $\m = 2$ bands, $E_{\pm}$,
separated by a gap from the rest of the spectrum. We also focus on the case of \emph{conical crossings}
\cite{Ha3}, which are, in a sense, the simplest non-trivial kind of crossings. The set of conical crossing
points $M_{\rm con} \subset \R^{3K}$ is generically a manifold of codimension $2$, see \cite{vNW}. Since
we are interested in the dynamics in the transverse direction we immediately assume $x \in \R^2$, $x=0$
being the only conical crossing point. The conical structure at crossing means that $E_{\pm}(x) = \pm
C_{\pm} |x|$ as $|x| \to 0$.

\medskip

The dynamics for the family $\{E_{+}, E_{-} \}$ is approximately described by an
effective Hamiltonian in the form (\ref{Effective m>1}). The choice of the basis
appearing in (\ref{U0 definition}) requires particular care near conical crossing
points. As pointed out already, away from the crossing manifold it is  convenient to
use the \emph{adiabatic basis}, \ie \, a basis $\{ \chi_a \}_{a= \pm}$ consisting of
eigenfunctions of the electronic Hamiltonian, see (\ref{EigenProb}). Such a basis is
uniquely defined, up to a choice of the phases, away from the crossing manifold.
Clearly, with this choice, $W(x)$ becomes a diagonal matrix. \noindent On the other
hand, $\nabla \chi(x)$ generically diverges as $x$ approaches a conical crossing,
and so does the Berry connection term, yielding an effective Hamiltonian with
singular coefficients. Diverging quantities are unstable when  numerical
discretization schemes are employed. Thus, if one is mainly interested in the
dynamics near the crossing points, it is advantageus to skip condition
(\ref{EigenProb}), and to work in a different basis, with the property that it
depends smoothly on $x$. Such a basis is called \emph{diabatic}.

\medskip

The next step is a reasonable truncation of the $\e$-expansion appearing in (\ref{Effective m>1}). We are
interested in the one-band Born-Oppenheimer dynamics near a conical crossing, where the functions $\A_+$,
$\phi_+$ and $\M_+$ are singular. Thus in (\ref{Effective m>1}) we retain the terms of order $\Or(1)$ and
between the terms of $\Or(\e)$ or higher we retain \emph{only the singular contributions}. Therefore, if
(\ref{Effective m>1}) is written in a diabatic basis, so that $\A(x)$ is a smooth function, we truncate
the expansion in (\ref{Effective m>1}) at the leading order, obtaining the model Hamiltonian $H_{\rm dia}=
\frac{1}{2} p^2 + W(x)$ (in a suitable diabatic basis). Viceversa, if the adiabatic basis is used, the
singular term $p \cdot \A + \A \cdot p$ has to be included.

The singular behavior of $\A_{+}$ at the conical crossing does not depend on the choice of the basis in
(\ref{U0 definition}). Indeed, in the adiabatic basis $W(x)$ is diagonal, and the \emph{one-band} Berry
connection for the upper band is given by $\A_{\rm adia, \, +} = \I \<\chi_{+}, \nabla \chi_{+} \>_{\Hf}$.
On the other side, in a generic basis $\{\ph_a\}_{a=1,2}$ one has a non-diagonal $\widetilde{W}(x)$ with
eigenvectors $\xi_{\pm}(x) \in \C^2$. In this basis the $\Or(\e)$ term contains
\begin{equation}\label{one-band Berry diabatic}
    \A_{\rm dia, \, +}(x) = \I \< \xi_+(x),\, \widetilde{\A}(x)\, \xi_{+}(x)\>_{\C^2} + \I \< \xi_{+}(x), \nabla
    \xi_{+}(x) \>_{\C^2},
\end{equation} where $\widetilde{\A}(x)_{ab} = \I \< \ph_a(x), \nabla_x \ph_b(x)\>_{\Hf}$. The first term
in (\ref{one-band Berry diabatic}) corresponds to the reduction from $H_{\rm e}$ to
the $2$-band Hamiltonian (\ref{Effective m>1}), while the second term is due to the
reduction to a specific energy band, namely $E_{+}$. Note that in the diabatic
representation the singular contributions come from the eigenvectors of $\tilde{W}$,
while the first term in (\ref{one-band Berry diabatic}) is smooth and can thus be
neglected for our purposes. The two basis are related by $\chi_a(x) = \sum_b
G_{ab}(x) \ph_{b}(x)$, with $G(x)$ an orthogonal matrix smoothly depending on $x$
outside the crossing point. With the help of (\ref{W tranforms}) and (\ref{A
transforms}) one easily checks that $ \A_{\rm dia, \, +}  = \A_{\rm adia, \, +}$. A
similar behavior is expected for the second order terms $\phi_{+}$ and $\M_{+}$,
although we are not in position to discuss this point here.

\medskip

 With the purpose of having an explicit example, we now focus on the model Hamiltonian
 $H_{\rm dia}= - \frac{1}{2} p^2  + W(x)$, acting in $L^2(\R^2, \C^2)$,
 with $x=(x_1, x_2)=(|x|\cos\varphi, |x|\sin\varphi)$ and
\begin{equation} \label{W caroline}
W(x) = C \( \begin{array}{cc}
            x_1  &  x_2 \\
            x_2  & -x_1
          \end{array}
\)  = C \,|x|  \(  \begin{array}{cc}
                   \cos \ph & \sin \ph \\
                    \sin\ph & -\cos \ph
                 \end{array}
       \), \qquad C > 0.
\end{equation} We freely switch between polar and cartesian coordinates in the following. The eigenvalues of
$W(x)$ are $ E_{\pm}(x) = \pm C \, |x| $ and a smooth family of eigenfunctions for $x \neq 0$ is given by
\begin{equation}\label{Caroline eigenvectors}
    \xi_{+}(x) = \E^{i \ph/2} \(\begin{array}{c}
                               \cos\frac{\ph}{2} \\
                               \sin\frac{\ph}{2}
                             \end{array}
   \) \qquad \qquad
   \xi_{-}(x) = \E^{i \ph/2} \(\begin{array}{c}
                               - \sin\frac{\ph}{2} \\
                                 \cos\frac{\ph}{2}
                             \end{array}
   \).
\end{equation} A direct computation yields
\begin{eqnarray}
  \nabla \xi_{+}(x) &=& \frac{1}{2|x|} \( \xi_{-}(x) + \I \xi_{+}(x) \) \mathrm{e}_{\ph}, \nonumber \\
  \nabla \xi_{-}(x) &=& \frac{1}{2|x|} \( - \xi_{+}(x) + \I \xi_{-}(x) \) \mathrm{e}_{\ph},
  \label{Nabla xi}
\end{eqnarray}
where $\mathrm{e}_{\ph} = |x|^{-1}(x_2, -x_1)$ in cartesian coordinates. The derivatives $\nabla
\xi_{\pm}$ are divergent at the crossing point. One easily computes the higher order Born Oppenheimer
approximation for the upper band, with the result
\begin{equation}\label{C1}
    \A_{+}(x) = \I  \< \xi_{+}(x), \nabla_x \xi_{+}(x) \> = - \frac{1}{2|x|}\mathrm{e}_{\ph},
\end{equation}
\begin{equation}\label{C2}
\phi_{+}(x) =  \< \nabla_x \xi_{+}, \cdot (1 - P_{+}(x)) \nabla_x \xi_{+} \> = \frac{1}{4|x|^2}.
\end{equation} Using
\[
\( W(x) - E_{+}(x) \)^{-1} \(1 - P_{+}(x)\) = - \frac{1}{2C|x|}\(1 - P_{+}(x)\)
\]
one obtains
\begin{equation}\label{C3}
\mathcal{M}_{+} = \ \frac{1}{2  C |x|} \sum_{i,j =1}^{2} p_i  \<  \d_i \xi_{+}(x), (1 - P_{+}(x)) \,
\d_j\xi_{+}(x) \>  p_j =  L(x,p) \frac{1}{4 C |x|^5} L(x,p),
\end{equation}
where $L(x,p) = x_1 p_2 - x_2 p_1$ is the angular momentum operator. Thus the effective Born-Oppenheimer
Hamiltonian to second order reads

\[
h_{+} = \frac{1}{2} \( p + \frac{\e}{2 |x|} \mathrm{e}_{\ph} \)^2 + C \, |x| + \e^2 \frac{1}{4|x|^2} +
\e^2 L(x,p) \frac{1}{4 C |x|^5} L(x,p) + \Or(\e^3).
\]\vspace{3 mm}

\noindent For the Hamiltonian of the lower band only  $\M_{-} = - \M_{+}$ changes, while $\A_{-} = \A_{+}$
and $\phi_{-} = \phi_{+}$. The inverse-square potential corresponds to the repulsive force of Berry and
Lim. On the other side, $\mathcal{M}$ contributes to the effective Hamiltonian with a more singular term,
whose sign depends on the electronic state: on the upper band it leads to a repulsive force, while on the
lower band it is of the same order, but attractive. In both cases it may dominate the contribution coming
from the Born-Huang potential.

It is instructive to reexpress $H_{\rm dia}$ in the basis (\ref{Caroline
eigenvectors}). Denoting by $S$ the unitary operator corresponding to the change of
basis, one gets
\[
H_{\rm adi} :=  S \, H_{\rm dia} \, S^{-1} = {\footnotesize \frac{1}{2}} \, p^2  +
\(\begin{array}{cc}
        E_+(x)  &    0     \\
        0       &  E_{-}(x) \\
\end{array} \)
 - \frac{\e}{|x|} \mathrm{e}_{\ph} \cdot p \( \begin{array}{cc}
  1 &  i \\
  -i & 1 \\
\end{array}\)   + \frac{\e^2}{2|x|^2}  \( \begin{array}{cc}
  1 &  i \\
  -i & 1 \\
\end{array}\).
\]
Since the transformation is unitary, no information as compared to $H_{\rm dia}$ has been discarded. If
for the dynamics in the upper band one merely restricts  $H_{\rm adi}$ to $a,b = +$, one recovers the
Born-Huang potential but misses the $\M $ correction.

So far we focused on the specific model (\ref{W caroline}). One may ask wether  the singularities computed
in (\ref{C1}), (\ref{C2}) and (\ref{C3}) are independent from the specific Hamiltonian considered. As for
the Berry connection $\A_{+}$, we already pointed out that the singularity does not depend on the choice
of the diabatic basis used to compute $W(x)$. \noindent Moreover, the sign of $\mathcal{M}$ does not
depend on the specific model. Indeed if two bands come very close, with $E_n(x) < E_{n+1}(x)$, but
themselves are well separated from the others bands, the main contribution to the reduced resolvent comes
from the $n$-th band. Thus
\begin{eqnarray*}
  - \( W(x) - E_{n+1}(x) \)^{-1}  P^{\perp}_{n+1}(x)  &=& -\sum_{m \neq n +1} \( E_m(x) - E_{n+1}(x) \)^{-1} P^{\perp}_{m}(x)  \nonumber\\
  & \approx & -\( E_{n}(x) - E_{n+1}(x) \)^{-1} P^{\perp}_{n}(x) > 0.
\end{eqnarray*}
Thus $\mathcal{M}_{n+1}$ is positive and, by the mirror argument, $\mathcal{M}_{n}$ is negative.


\subsection{Second order corrections in chemical reactions}

The search for observable effects of the Berry phase has been a wide field of
investigations \cite{BMKNZ}. In the same spirit, one may ask if there are observable
effects related to the second order corrections $\phi$ and $\M$. A good candidate is
the chemical exchange reaction involving a system of three hydrogen atoms: $H_2 + H
\rightarrow H + H_2$. Indeed, the intermediate transition molecule $H_3$ exhibits an
interesting structure: two electronic energy bands form a family of conical
intersections, corresponding to those highly-symmetric configurations where the
nuclei are at the vertices of an equilateral triangle.

Experimental studies of $H_2 + H$ (which are usually performed by using one of its isotopic analogues, as
$ D + H_2 \rightarrow DH + H$, so that the reactants are labelled) have reached a very high level of
precision. For example, in \cite{experiment} the rates are measured for reactions in which both the
initial and the final hydrogen molecules are in a specific vibrational and rotational state. In fact, one
expects  the state-to-state cross section to be more sensitive than the total cross section to phase
interference of the Born-Oppenheimer wave function.

The basic interference mechanism is simple. The labelled reaction $A + BC \rightarrow AB + C$ may happen
in two qualitatively  different ways: either (i) the incoming atom $A$ binds directly with $B$, or (ii)
atom $A$ first approaches atom $C$ before finally binding with atom $B$. In the Born-Oppenheimer
approximation, this two possibilities correspond to classical paths on opposite sides of the conical
crossing. Thus quantum interference is expected.

On the theoretical side, Kupperman and Wu \cite{KuWu} computed state-to-state differential cross section
for the $H + H_2$ reaction, showing significant difference with previous calculations, where the Berry
phase was neglected. Comparison between theory and experiments was firstly done for $D +H_2 \rightarrow DH
+ H$. In early computations, the inclusion of the Berry phase improved significantly the agreement between
theory and experiments. However, it has been later pointed out, the agreement could be due to accidental
cancellations of errors, one source of error being the inaccuracy of the electronic structure data in the
high-energy region. The results are still controversial, and no general consensus has been reached
\cite{BMKNZ}.

\bigskip

While we do not enter in the controversy, we comment on the relevance of
second-order Born-Oppenheimer corrections in this kind of reactions.  First of all,
some orders of magnitude: in the identical atoms system the conical intersection
exists at $E_{\rm cross} \approx 2,7\ eV$ above the minimal energy of the $H_2$
molecule, and the minimal potential barrier is $ E_0 \approx 0.43\ eV$,
corresponding to the collinear $H-H-H$ configuration.  We focus on a situation in
which the kinetic energy is smaller than $E_{\rm cross}$, so that diabatic
transitions to the upper band are negligible, but sufficiently larger that $E_0$, so
that the probability that the wavefunction follows a path of kind (ii) is not
negligible. In this energy interval the wavefunction is likely to explore regions of
the nucleonic configuration space close to the crossing point. The second order
corrections $\phi$ and $\M$ are singular at the crossing point, and thus certainly
relevant in the nearby region. Although  a careful analysis requires quantitative
estimates, we conclude that the second-order terms might be relevant in this kind of
reactions.


\end{document}